\documentclass[fleqn,10pt]{wlscirep}
\usepackage[utf8]{inputenc}
\usepackage[T1]{fontenc}

\newcommand \MZ [1]{\bgroup\noindent[\textcolor{blue}{\textbf{MZ}: #1}]\egroup\ignorespacesafterend}

\title{Tuning load redistribution and damage near heterogeneous interfaces}

\author[1]{Christian Greff}
\author[1,*]{Paolo Moretti}
\author[1]{Michael Zaiser}
\affil[1]{Dept. of Materials Science, WW8-Materials Simulation, FAU Universit\"at Erlangen-N\"urnberg, Dr.-Mack-Stra{\ss}e 77, 90762 F\"urth, Germany}
\affil[*]{paolo.moretti@fau.de}

\begin{abstract}
We investigate interface failure of model materials representing  architected thin films in contact with heterogeneous substrates. We find that, while systems with statistically isotropic distributions of impurities derive their fracture strength from the ability to develop rough detachment fronts, materials with hierarchical microstructures confine failure near a prescribed surface, where crack growth is arrested and crack surface correlations are suppressed. We develop a theory of network Green's functions for the systems at hand, and we find that the ability of hierarchical microstructures to control failure mode and locations comes at no performance cost in terms of peak stress and specific work of failure and derives from the quenched local anistotropy of the elastic interaction kernel.  
\end{abstract}
\begin{document}

\flushbottom
\maketitle
%
%
\thispagestyle{empty}

\section*{Introduction}

The role of microstructural randomness in fracture and failure has been variously emphasized in the past, noting the importance of fluctuations in local strength and density for the roughening of crack fronts and the emergence of rough fracture surfaces \cite{Bonamy2006_PRL,alava2006statistical,alava2008role}. As opposed to the case of homogeneous specimens, where failure is induced by the expansion of smooth cracks, rough crack profiles in heterogeneous materials point to a slower nucleation-plus-propagation scenario, where quenched disorder deflects the advancing crack front and delays failure. In recent years, hierarchically architectured microstructures have been explored as another approach to impede crack propagation by inducing crack deflection and arrest \cite{Lakes1993,fratzl2007nature,gao2006application}. In hierarchcally architectured materials, elastic load propagation in a hard and brittle matrix is modulated by the presence of  soft and compliant lamellar inclusions or void-like planar gaps \cite{Sen2011,mirzaeifar2015defect,Moretti2018}, which interrupt stress transmission an facilitate crack deflection. Such hierarchical microstructures are inspired by certain examples of flaw-tolerant protein-based biological materials, such as collagen and spider silk, where fibrous patterns repeat across scales in a self-similar fashion \cite{Sun2012, Jiao2015, Gao2006, Rho1998, Gautieri2011, Sen2011,roemer2008prion,lu2023silk}. Numerical and experimental results confirm the idea that hierarchical patterning efficiently mitigates against stress/strain concentrations at crack tips, effectively arresting propagation and promoting a diffuse mode of failure that is insensitive to the existence of even large flaws \cite{Zaiser2022,pournajar2023failure} and that has been recently associated with the formation of multifractal fracture surfaces \cite{Hosseini2021,hosseini2023enhanced}.

The mechanical advantages of hierarchical microstructures have also been observed in the context of thin-film interface failure\cite{Puglisi2013_PRE,Esfandiary2022} and  friction \cite{Costagliola2016_PRE,Costagliola2022_IJSS}, mostly inspired by the Gecko foot as a paradigm of smart, recyclable adhesives \cite{KimJAST_2007,Sauer2014_JAST,Bhushan2009_PTRS,Gao2005_MM}. In analogy with the case of bulk fracture, numerical models of interface failure of hierarchically patterned network structures confirm the appearance of crack arrest phenomena, although different geometrical factors come into play. First, the thin film geometry imposes an effective cutoff to the otherwise long-ranged elastic kernel of a statistically homogeneous medium, to the extent that, on scales above the film thickness, load re-distribution becomes effectively local \cite{fyffe2004effects,Barai2013_PRE,Taloni2015_SciRep}. This raises the question to which extent hierarchical patterning can induce crack arrest by modulating stress transmission. Second, in interface failure random heterogeneity may be overruled by the intrinsic heterogeneity of the thin film system, most notably because the adhesive forces between film and interface are often inherently weaker than cohesive forces (with some exceptions \cite{yao2024mechanical}), and this may tend to localize failure at the interface and thus prevent roughening of the fracture surface. Again this observation seems to point to a reduced 
importance of the thin film microstructure away from the interface. 

Statistical analysis of numerically simulated fracture surfaces reinforces these fundamental observations. Non-hierarchical network models of fracture have long shown that in bulk systems, strength heterogeneity results in self-affine fracture surfaces \cite{Bonamy2006_PRL,alava2006statistical,alava2008role}. By contrast, simulations of non-hierarchical thin film geometries \cite{Barai2013_PRE} indicate fracture surfaces with multiscaling properties. The same is true for hierarchical 3D bulk systems \cite{hosseini2023enhanced}. A conclusive multiscaling analysis of systems displaying both thin film geometries and hierarchical microstructures is still lacking. 

Here we address the theoretical and numerical modeling of interface failure of microstructured  thin films, where microstructural patterns are induced by the simple process of matter removal. Our structures are subjected to uniaxial tensile loads, thus mimicking a peeling phenomenon.  In particular we consider two model materials, namely i) heterogeneous but statistically isotropic systems, where removal is implemented as random micro-void formation; and ii) hierarchical systems, where removal consists of planar cuts or gaps of variable depth, originating at the interface, and inspired by the discontinuities in the fibrous structure of gecko foot contact surfaces. Materials microstructures and local mechanical response are described using network models, by representing a material as a network of load-carrying, breakable edges. Such models have been successfully introduced in recent years for hierarchical materials, in order to study the relationship between their microstructure and failure behavior. The local mechanical material response is simulated here using the Random Fuse Model (RFM) \cite{deArcangelis1985_JPL}. While more sophisticated approaches exist, where a full description of edges as Timoshenko beams is used and Maximum Distortion Energy arguments are put forward to develop rules for single-beam failure  \cite{Hosseini2021,hosseini2023enhanced}, simpler scalar models such the RFM become relevant in large-scale statistical studies like ours, where the computational cost of beam simulations would be significant. 

Our results show that, in the context of adhesion and detachment, hierarchical structures allow for control over both failure location and failure mode. In terms of failure location, hierarchical microstructures ensure that detachment occurs at the prescribed interface, even in extreme situations of comparatively high adhesion forces. This degree of control is achieved by promoting a failure mechanism that is mostly confined at the interface, resulting in fracture surfaces that exhibit only small fluctuations with very short-range correlations, and detachment being the result of damage accumulation rather than crack growth. In order to investigate the mechanisms of stress redistribution in these systems, we introduce the concept of a fuse network Green's function, by transferring early graph theory results on random walks and Markov processes \cite{chung2000_green} to the materials problem at hand. Both non-hierarchical thin films and their hierarchical counterparts display a screened form of elastic load re-distribution, with interaction kernels exhibiting exponential cutoffs controlled by the film thickness. However, in hierarchical systems this kernel exhibits pronounced local anisotropies which average out only globally, but not over the effective kernel range. This leads to local directional modulations of stress transmission which may trigger crack arrest.

\section*{Methods}

\subsection*{Microstructure models}

\begin{figure}[t]
\centering
\includegraphics[width=0.9\linewidth]{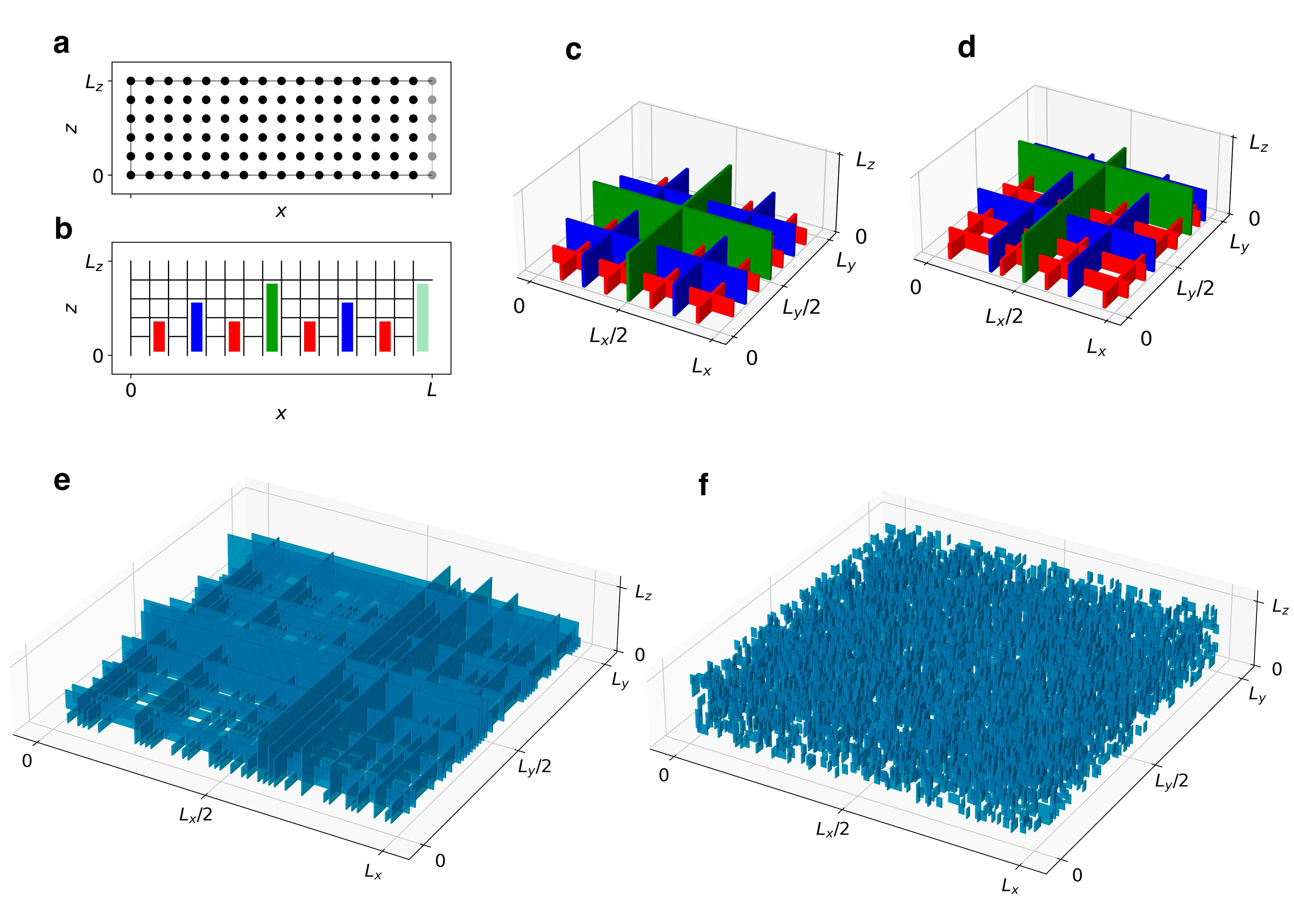}
\caption{Network models. (a) In a system of $s$ hierarchical levels, $N = L^2 (L_z+1)$, nodes are distributed in a 3D cubic lattice of lateral size $L=2^s$ and thickness $L_z=1+s$. The case of $s=4$ is shown. The boundaries at $x=L$ (shown in gray) and $y=L$ (not shown) are periodic. The nodes are connected by edges to form a fully connected cubic lattice; (b) Horizontal (film-parallel) edges are recursively removed to form cuts which create a deterministic hierarchical cut pattern. The $s$-th hierarchical level comprises no cuts. For $0<i<s$, the  $i$-th hierarchical level comprises $2^{s-i-1}$ cuts of height $i$ in the $x$ and an equal number in the $y$ direction. The cases $i=1$, $i=2$, $i=3$ are shown in red, blue, green respectively. Periodic boundaries  are introduced as additional highest-level cuts (pale green). (c) 3D view of the deterministic cut structure of (b). For clarity of representation, the two pale green boundary cuts are not shown. (d) A single realization of the hierarchical network model for $s=4$ obtained as a stochastic variant of (b). (e) Single realization of the hierarchical network model for $s=6$, corresponding to $L=64$, the smallest system size considered in this work. (f) Non-hierarchical variant of (e), where removed horizontal edges are randomly distributed throughout the system.}
\label{fig:model}
\end{figure}

We consider adhesion systems modeled as thin films in contact with a rigid substrate. The film is envisaged as a box of linear size $L_x=L_y=L$ and $L_z\ll L$, and the frame of reference is chosen so that the film extends in the interval 
$[0,L]$ along the $x$ and $y$ directions, and $[0,L_z]$ in the $z$ direction. Boundaries at $x=L$ and $y=L$ are periodic, while $z=0$ identifies the contact surface (Figure \ref{fig:model}). To model hierarchical (and non-hierarchical) microstructural arrangements, dimensions are chosen such that, given a positive integer $s$, $L=2^s$ and $L_z=1+s$.  We introduce a network discretization of the film, where $N = L^2(L_z+1)$ nodes are placed at the sites with at integer values of $x$, $y$, and $z$ (Figure \ref{fig:model}a). Starting from this node layout, edges are added as a subset of all the possible edges in a perfect cubic lattice (Figure \ref{fig:model}b). While all edges along $z$ are retained, edges along $x$ and $y$ are selectively removed depending on the distance from the interface in an $s$-step process, in order to mimic hierarchical contact for a system comprising $s$ hierarchical levels. Edge removal introduces cuts (or gaps) in the system, parallel to either the $xz$ or the $yz$ planes, spanning the whole system in the $xy$ plane, and displaying the whole range of heights from 1 to $L_z-1$. In a preliminary step, a cut pattern is implemented following the deterministic method first introduced by Esfandiary {\it et al.}\cite{Esfandiary2022} (Figure \ref{fig:model}b-c). Then, this deterministic hierarchical pattern is used to build an ensemble of stochastic hierarchical structures by randomly reassigning the position of each gap in the $xy$ plane (Figure \ref{fig:model}d-e). Our numerical simulations on hierarchical systems run on multiple realizations of this stochastic variant. An ensemble of reference, non-hierarchical systems is also constructed, by randomly reassigning the position of each individual removed horizontal edge, thus eliminating the gap organization and resulting in a statistically homogeneous distributions of voids (Figure \ref{fig:model}f).

\subsection*{Elasticity and network Green's function}
To model adhesion and detachment properties, we consider the simple case in which the system is already in contact with the interface ($z=0$) and is being peeled by a force acting on the top boundary ($z=L_z$). We model the elastic response of the system and its failure in the RFM framework. 
The $N$ nodes are labelled using indices $i\in[0,N-1]$ and the connectivity information is stored in the adjacency matrix $\mathbb{A}$, whose generic element $A_{ij}$ is $1$ if nodes $i$ and $j$ are  connected by an edge, and $0$ otherwise. 
We call $\Omega$ the set of all nodes, and $\partial\Omega$ its boundary, consisting of all nodes at $z=0$ and $z=L_z$. In particular, we distinguish the set of bottom and top boundary nodes as $\partial\Omega_0$ and $\partial\Omega_{L_z}$, with $\partial\Omega=\partial\Omega_0 \cup\partial\Omega_{L_z}$
Each node $i$ has a displacement-like variable $u_i$ and each edge $ij$ between nodes $i$ and $j$ carries a force-like variable $f_{ij}$. Since elastic variables are envisaged as scalars, elastic behavior is enforced by a scalar Hooke's law of the form 
\begin{equation}
\frac{f_{ij}}{X}=E \frac{u_i-u_j}{\ell}
\end{equation}
where $X$ and $\ell$ are the edge cross-section and length respectively, and the elastic modulus $E$ is the proportionality constant between fuse stress $f_{ij}/X$ and fuse strain $(u_i-u_j)/\ell$. In the following, we choose $\ell$ as the unit of length and, for convenience of notation, we introduce the fuse stiffness $\kappa = EX/\ell$,
in a form that is reminiscent of e.g. axial stiffness in beam elasticity. Elastic equilibrium at each node $i$ is imposed through the system of algebraic equations
\begin{equation}\label{eq:discrete_laplace}
\sum_{j\in \Omega} L_{ij} u_j = \frac{1}{\kappa}f^*_i,
\end{equation}
where we have introduced the discrete Laplace operator $\mathbb{L}$ with matrix elements $L_{ij}$. For the generic node $i$,   $L_{ij}=\delta_{ij}\sum_{l\in \Omega}A_{jl}-A_{ij}$
(graph Laplacian matrix), and $f^*_i$ is the sum of external body forces acting on the node $i$. Solutions of Equation \ref{eq:discrete_laplace} depend on the choice of boundary conditions. External loads in our uni-axial geometry can be applied in the form of displacements, e.g. applying $u_i=0$ at the lower boundary $i\in\partial\Omega_0$ and $u_i=U>0$ at the upper boundary $i\in\partial\Omega_{L_z}$. The equilibrium equations thus become
\begin{equation}\label{eq:dirichlet}
\sum_{j\in\Omega\setminus\partial\Omega} L_{ij}u_j=b_i
\;\;\;\;\;\;\mathrm{with}
\;\;\;\;\;\; i\in\Omega\setminus\partial\Omega
\end{equation}
where indices $i$ and $j$ extend to non-boundary nodes and the vector of elements $b_i$ is known and contains the body forces on $i$ as well as the forces exerted by neighboring boundary nodes, namely $b_i=\sum_{j\in\partial\Omega}A_{ij}u_j
+f^*_i/\kappa$. 
Equation \ref{eq:dirichlet} defines a non-singular submatrix $\tilde{\mathbb{L}}$ of $\mathbb{L}$, obtained by removing the $B$ rows and columns corresponding to boundary nodes \cite{Moretti2019_EPJB}. Introducing  $\mathbb{G}=\tilde{\mathbb{L}}^{-1}$, the algebraic problem thus has $N-B$ unknowns and admits a unique solution, the elastic equilibrium state $u_i$
\begin{equation}\label{eq:discrete_solution}
u_i = \sum_{j\in\Omega\setminus\partial\Omega}G_{ij}b_j
\;\;\;\;\;\;\mathrm{with}
\;\;\;\;\;\; i\in\Omega\setminus\partial\Omega ,
\end{equation} or in vector form
$
\mathbf{u} = \mathbb{G}\mathbf{b}
$,
where vectors are defined in the $N-B$-dimensional space.  In our RFM simulations, Equation \ref{eq:discrete_solution} is solved numerically, supplying $\tilde{\mathbb{L}}$ and $\mathbf{b}$ to a parallel sparse solver. In particular, in simulations we consider the Dirichlet problem with uniform displacement $U>0$ at the top layer and no body forces, $f_i^*=0$. For our analytical work, instead, let us consider the case of vanishing displacement at the boundaries and non-zero applied body forces, so that $b_i=f_i^*$. It becomes clear that in \ref{eq:discrete_solution}, the matrix $\mathbb{G}$ acts as the network equivalent of the elastic Green's function in the continuum problem. We call $\mathbb{G}$ the network elastic Green's function, in analogy with earlier works in spectral graph theory, where a similar quantity was introduced in the context of random walks \cite{chung2000_green}. While in random walks, $\mathbb{G}$ is the propagator of a stochastic process, in our elastic problem  $\mathbb{G}$ is the discrete kernel of elastic interactions. In particular, it easy to see that $G_{ij}$ is the displacement at node $i$, as produced by a unit point force at $j$, equal to $f_j^*=1\times \ell/\kappa$. 

\subsection*{Elastic Green's function in the continuum limit}
For an isotropic system, the continuum limit of Equation \ref{eq:dirichlet} is obtained with the usual transformation
\begin{equation}
\ell^{-2}L\to -\nabla^{2}.
\end{equation}
We can thus rewrite Equation \ref{eq:dirichlet} as a Poisson equation of the form 
$-\nabla^2 u(\mathbf{r})=b(\mathbf{r})$, with corresponding Green's function defined by
\begin{equation}
-\nabla^2 G(\mathbf{r},\mathbf{r'}) = \delta(\mathbf{r}-\mathbf{r'}).
\end{equation}
In this paper, we compute $G(\mathbf{r},\mathbf{r'})$ using 
Fredholm's theory. Choosing an appropriate basis of eigenfunctions $\varphi_{l,m,n}(\mathbf{r})$ of $-\nabla^2$, with corresponding eigenvalues $\lambda_{l,m,n}$, the Green's function is given by
\begin{equation}\label{eq:fredholm}
G(\mathbf{r},\mathbf{r'})=\sum_{l,m,n} \frac{\varphi^*_{l,m,n}(\mathbf{r'})\varphi_{l,m,n}(\mathbf{r})}{\lambda_{l,m,n}}.
\end{equation}
Once the Green's function is known, the solution to the inhomogeneous Poisson equation is 
\begin{equation}\label{eq:continuum_solution}
u(\mathbf{r})=\int d^3r'G(\mathbf{r},\mathbf{r'})b(\mathbf{r'}),
\end{equation}
which is the continuum version of \ref{eq:discrete_solution}. In order to enforce the thin-film geometry with $L_z\ll L$, we choose $L\to\infty$ limit, so that $\varphi_{l,m,n}(\mathbf{r}) \to \varphi_{k_x,k_y;\,n}(\mathbf{r})$, where $k_x$ and $k_y$ are continuous variables. An appropriate basis of eigenfunctions 
$ \varphi_{k_x,k_y;\,n}$, which describes the system in the $L\to\infty$ limit, and vanish at $z=0$ and $z=L_z$, is the set of eigenfunctions for the wave equation \cite{Jackson_book}, namely
$
 \varphi_{k_x,k_y;\,n} = \psi_{k_x}(x)\psi_{k_y}(y)\chi_{n}(z),
$
with 
\begin{equation}
\psi_{k_x}(x)=\frac{1}{\sqrt{2\pi}}\mathrm{e}^{\mathrm{i}k_x x},\;\;\;\;\;\;
\psi_{k_y}(y)=\frac{1}{\sqrt{2\pi}}\mathrm{e}^{\mathrm{i}k_y y},
\end{equation}
and 
\begin{equation}
\chi_{n}(z)=\sqrt{\frac{2}{L}}\sin k_n z
\;\;\;\;\;\;\mathrm{with}\;\;\;\;\;\;
k_n = \frac{\pi n}{L_z}
\;\;\;\;\;\;\mathrm{and}\;\;\;\;\;\;
n=1,2,3\dots. 
\end{equation} 
We note that the eigenvalue of $-\nabla^2$, corresponding to the generic eigenfuction $\varphi_{k_x,k_y;\,n}$ simply is $k_x^2+k_y^2+k_n^2$, so that the Green's function is
\begin{equation}
G(x-x',y-y';z,z')=\sum_{n=1}^{\infty}
\int\frac{dk_x}{2\pi}\int\frac{dk_x}{2\pi}
\mathrm{e}^{\mathrm{i}\left[k_x(x-x')+k_y(y-y')\right]}
\frac{\chi_{n}(z)\chi_{n}(z')}{k_x^2+k_y^2+k_n^2}.
\end{equation}
Since the angular dependence in $k_x,k_y$ is contained only in the exponential terms, the double integral describes a standard Hankel transform, which can be computed as 
\begin{equation}\label{eq:green_solution}
G(x-x',y-y';z,z')=\frac{2}{L_z}\sum_{n=1}^{\infty}
\sin\left( \frac{\pi n}{L_z}z \right)
\sin\left( \frac{\pi n}{L_z}z'\right) 
K_0\left[\frac{\pi n}{L_z}r \right],
\end{equation}
where $r=\sqrt{(x-x')^2+(y-y')^2}$, and $K_0(\dots)$ is a modified Bessel function of the second kind. Equation \ref{eq:green_solution} is the exact Green's function for the isotropic system in the $L\to\infty$ limit, and describes the scalar-elastic displacement field at position $(x,y,z)$ as produced by a Dirac delta perturbation at $(x',y',z')$. As we are interested in screening effects at distances $r> L_z$, we note that $K_0$ admits the asymptotic expansion $K_0(r)=\sqrt{\pi/2}r^{-1/2}\mathrm{e}^{-r}\left[1-\mathcal{O}\left(1/r\right) \right]$
so that at the lowest order we can finally estimate
\begin{equation}\label{eq:first_order}
G(x-x',y-y';z,z') \approx \frac{\sqrt{2\pi}}{L_z}
\sin\left( \pi\frac{z}{L_z} \right)
\sin\left( \pi\frac{z'}{L_z} \right)
\left(\pi\frac{r}{L_z}\right)^{-\frac{1}{2}}
\mathrm{e}^{-\pi\frac{r}{L_z}}.
\end{equation}

\subsection*{Failure criterion and simulation protocol}

On the edge level we consider ideal elastic-brittle behavior of the individual edges $ij$. Thus, the edges behave elastic until the force reaches a force threshold $t_{ij}$, and is irreversibly removed as soon as the evaluated force reaches levels $f_{ij}>t_{ij}$ \cite{alava2006statistical}. Each network realization is characterized by a set of $t_{ij}$, extracted from a given threshold distribution. Different distributions can be used in order to parametrize heterogeneity in the strength of the constituents of a system. In particular, we resort to Weibull distributions with cumulative distribution function 
\begin{equation}
C(t)=1-\mathrm{e}^{-\left(\frac{t}{\lambda}\right)^k},
\end{equation}
where $\lambda$ is the scale parameter and $\beta$ the shape parameter. By imposing that $C(t)$ has mean value $\bar{t}$, we rewrite the scale parameter as $\lambda=\bar{t}/\Gamma(1+1/k)$. With this choice of parameters we can tune the threshold distribution by acting on $\bar{t}$, which equals the average local strength, and $k$, which controls the degree of heterogeneity, with larger $k$ corresponding to narrower distributions and lesser fluctuations. 

\begin{figure}[tbhp]
\centering
\includegraphics[width=0.8\linewidth]{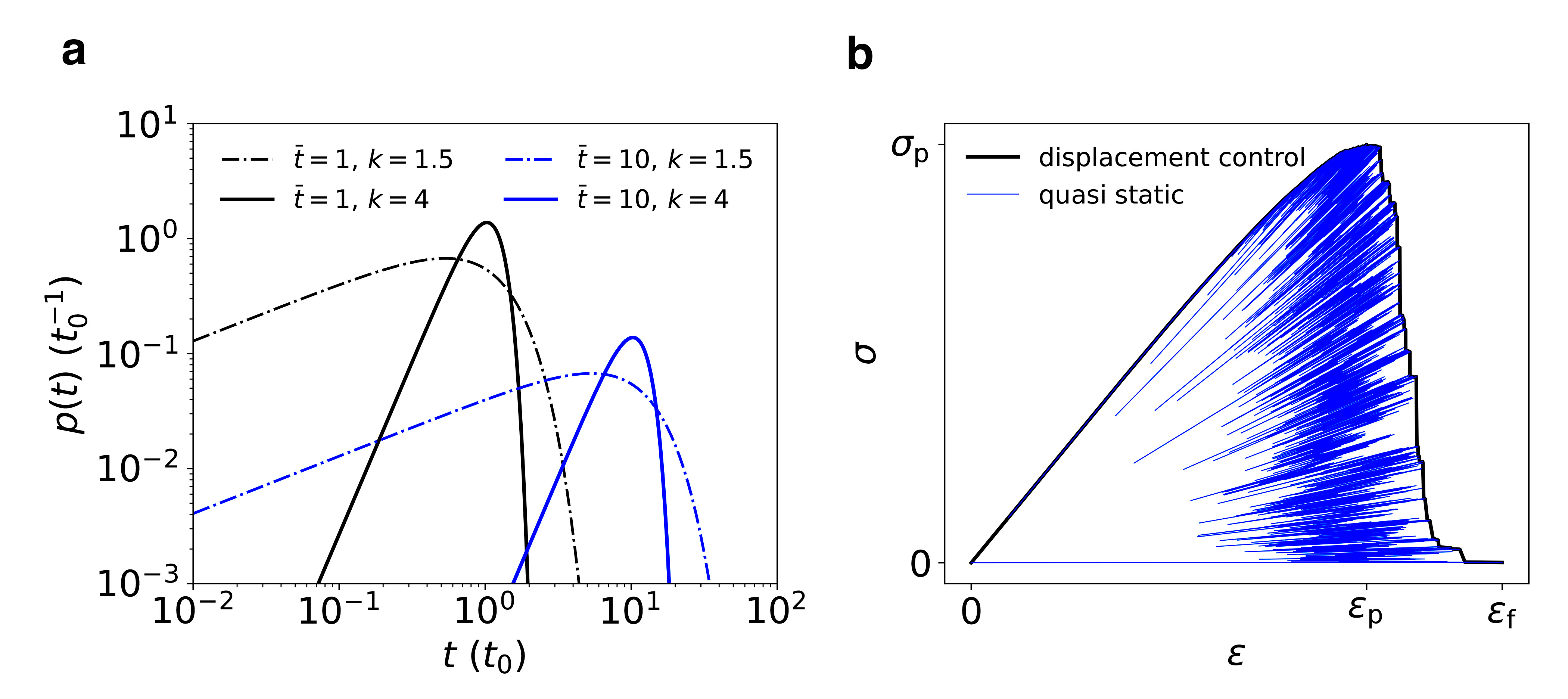}
\caption{Simulation details; (a) probability density functions of the failure thresholds used in this work. High heterogeneity: $k=1.5$ (dash-dotted lines). Low heterogeneity: $k=4$ (solid lines). Distributions with average $\bar{t}=10$ (blue lines) describe situations where cohesive forces are much larger than adhesive forces; (b) typical stress-strain curve under displacement control (thick black line), constructed by enveloping the stress-strain curve obtained from a quasi-static simulation (thin blue line).}
\label{fig:details}
\end{figure}

In our simulations we tune the parameters $\bar{t}$ and $k$ in order to explore limiting cases of relevance to the interface failure problem (Figure \ref{fig:details}a). In regard to average thresholds, we consider two opposing scenarios: (a) $\bar{t_{ij}}=t_0$ for every fuse, i.e. local strengths are on average the same everywhere in the system; (b) $\bar{t_{ij}}=t_0$ for edges emanating from the interface (lower boundary) and $\bar{t_{ij}}=m\,t_0$ ($m>1$) elsewhere, modeling a more realistic situation in which cohesive forces are larger than adhesive forces, resulting in a \textit{weak layer} at the interface. Without loss of generality we choose $t_0=1$, so that we measure forces in units of $t_0$ and stresses in units of $\sigma_0=t_0/X$. For the weak interface case, we choose $m=10$, pointing to a scenario in which cohesive forces are one order of magnitude stronger than adhesive forces. Larger values of $m$ would produce results that tend to those of previous studies, which have explored the simple case of $m\to\infty$ \cite{Esfandiary2022,Barai2013_PRE}. 

The global stress is computed starting from the global force $F$ acting on a cross section of the system as $\sigma = F/(L_xL_y)$. Similarly, the global strain is computed starting from the boundary displacement $U$ as $\epsilon = U/L_z$;  it is measured in units of $\sigma_0/E$.

In order to model the response to uniaxial loads, we determine averages of the peak stress $\sigma_\mathrm{p}$ and  work of failure $W_\mathrm{f}$, for both notched and un-notched systems, in displacement control. To this end we choose the standard quasi-static simulation procedure, where a constant displacement $U$ is applied \cite{alava2006statistical}. At every step, Equations \ref{eq:dirichlet} are solved for every non-boundary node $i$, edge-wise forces are computed, the edge with the maximum load factor $\rho_{\rm max}:={\rm max}_{i,j}(f_{ij}/t_{ij})$ is identified as the weakest and removed. The local and global forces and displacements are rescaled by the factor $\rho_{\rm max}$, thus effectively setting the displacement to the precise value when the weakest edge fails. While this method describes an idealized protocol, where at every step displacements are adjusted to the minimum values allowing for the removal of the weakest link, the desired protocol in which displacement is monotonically increased can be obtained by enveloping the resulting stress-strain curves. This procedure is exemplified in Figure \ref{fig:details}b. For an individual run, the peak stress $\sigma_\mathrm{p}$ is the maximum $\sigma$, which points to the maximum load that a structure can carry and provides an adequate strength descriptor for very brittle materials with little post-peak resistance.
The work of failure $W_\mathrm{f}$ (per unit volume) is the area under the stress-strain curve, and  quantifies the energy per unit volume that is necessary to reach failure, including the post-peak regime.


\subsection*{Surface analysis}
We define the fracture surface as the spanning cluster of failed edges that emerges when the load reaches zero and the sample loses vertical connectivity. The height $h_i=h(x_i,y_i)$ of the crack profile is then defined as the minimum $z$ distance between the fracture surface and the node $i$ in the $z=0$ plane. We call $p(h)$ the probability of observing crack height $h$ when averaging over an ensemble of realizations of a system with the same choice of parameters (type, size, threshold distributions). In order to study the scaling properties of fracture surfaces, we measure the structure function \cite{Barabasi1991_PRA,Barabasi1992_PRA,Picallo2009_PRL,Barai2013_PRE}
\begin{equation}\label{eq:structure_defintion}
F_q(r)=\langle[ \langle \left|h-h' \right|^q \rangle_{r} \rangle]^{\frac{1}{q}},
\end{equation}
where the average $\langle \dots \rangle_{r}$ runs over pairs of nodes in the $z=0$ plane that are at Euclidean distance $d=r$. 
For surfaces exhibiting ideal self-affine scaling behavior, $F_q(r)\sim r^H$ for every $q$ and for all $r$, where the constant $H$ is referred to the Hurst or roughness exponent. In practical terms, we may speak of self affine scaling when $q$ is approximately constant at least within a finite range of positive $q$ and for a scaling range of $r$ values that spans at least an order of magnitude. 

In case of self affine scaling, the value of $H$ is equal to the exponent controlling the scaling of the so called local width $w(r)$, measuring the standard deviation of $h$ within a window of size $r$\cite{bustingorry2021_JoP}: $w(r) \propto r^H$. Self affine scaling breaks down if $H_q$ depends on $q$, when the system is said to display multiscaling behavior. In this case, no simple scaling laws can be put forward to relate fluctuations, correlations and other collective observables. Finally, we also compute the correlation function
\begin{equation}\label{eq:correlation_definition}
C(r) = \langle hh' \rangle_{r} - \langle h\rangle \langle h' \rangle,
\end{equation}
which, unlike $F_q(r)$, decreases with $r$, recording the loss of height correlations at large distances.  We note that in the literature the function $[F_q(r)]^q$ is sometimes referred to as height-height correlation function, however we avoid to use this name, in order to avoid terminological confusion with the correlation function $C(r)$. In particular, no simple relationship exists between the correlation and structure functions whenever the scaling hypothesis is not verified.

Given the periodicity of our systems in the $x$ and $y$ directions, in all cases $F_q(r)$ and $C(r)$ are computed within square domains of linear size $L/2$, centered at each node $i$, and averaged over all choices of $i$ and all realizations.

\section*{Results}

\subsection*{Crack localization}
\begin{figure}[t]
\centering
\includegraphics[width=\linewidth]{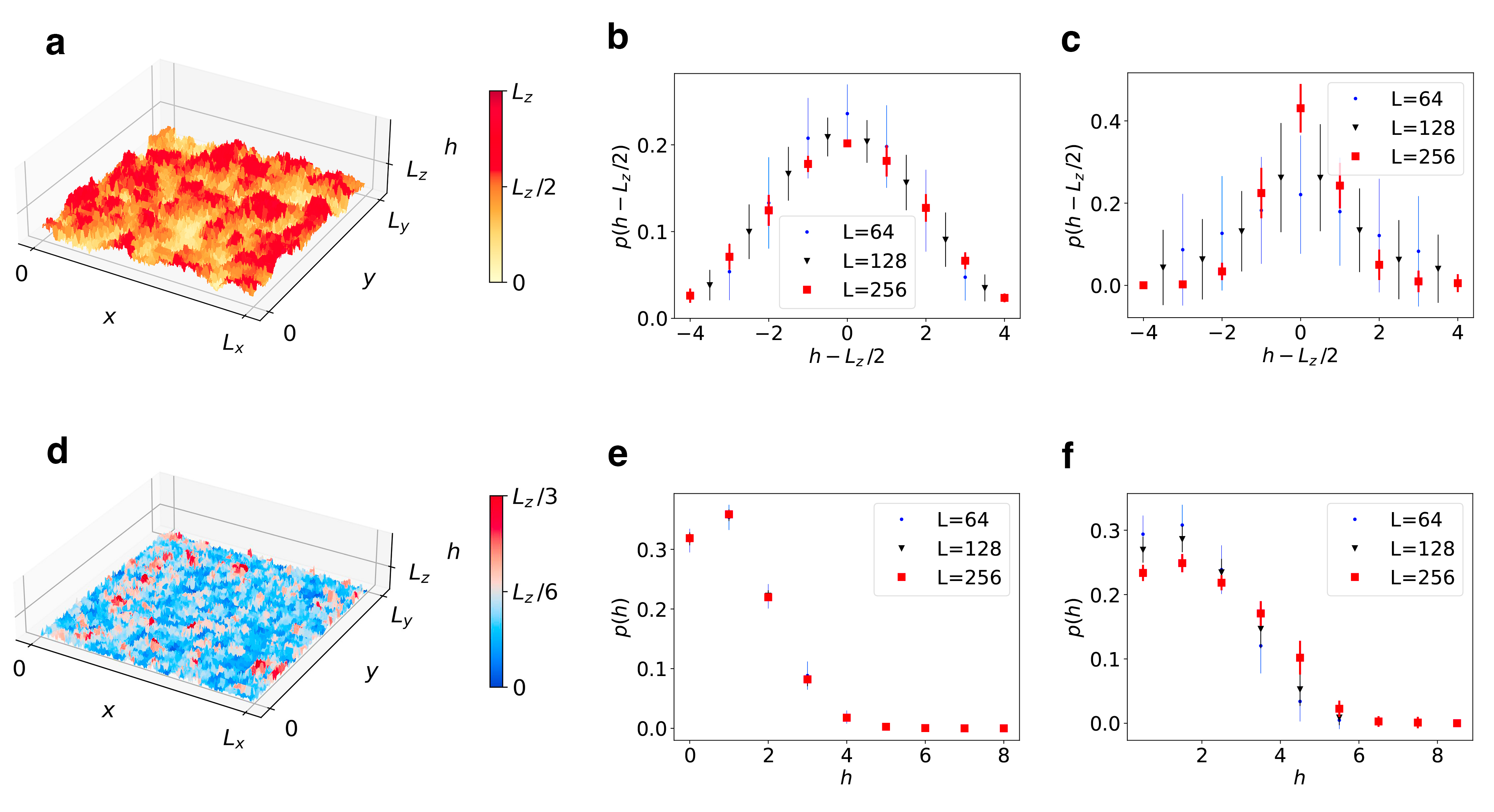}
\caption{Fracture profiles. (a) Fracture height profile $h=h(x,y)$ for a non-hierarchical sample, $L=256$, $k=1.5$. (b) and (c) Probability density $p(h)$ for non-hierarchical systems, $k=1.5$ and $k=4$ respectively. 
(d) Fracture height profile $h=h(x,y)$ for a hierarchical sample, $L=256$, $k=1.5$. (e) and (f) Probability density $p(h)$ for hierarchical systems, $k=1.5$ and $k=4$ respectively. 
}
\label{fig:profiles}
\end{figure}

In order to evaluate the ability of hierarchical systems to localize fracture and detachment surfaces at the interface, we first performed simulations for both hierarchical and non-hierarchical systems of multiple sizes, under the assumptions that no notches are present at the interface and the average of local threshold distributions equals $t_0$ everywhere (no weak layer, $m=1$) such that the interface is, from the viewpoint of strength, equivalent to the bulk material. Our results are summarized in Figure \ref{fig:profiles}, where we analyze the heights of crack surfaces $h(x,y)$. As expected of heterogeneous, non-hierarchical materials, our reference model displays rough crack surfaces, with crack height probability densities that are statistically symmetric around $h=L_z/2$ where they exhibit a maximum (Figure \ref{fig:profiles}a-c). Interestingly, the limited thickness $L_z$ ensures that an extended region characterized by bulk behavior is never observed, and profiles are heavily influenced by boundary effects; the mean location of the fracture surface is in the central plane of the film and does nowhere extend to the top or bottom surface of the film. 

In hierarchical systems, results are radically different (Figure \ref{fig:profiles}d-f), as detachment surfaces are indeed localized at the interface, as an indirect consequence of the fact that all cuts emanate from $z=0$. The network structure thus allows us to prescribe the fracture location, even without introducing weak layers with lower failure strengths. Non-hierarchical and more in general statistically isotropic systems derive their fracture strength from their ability to store damage in rough cracks. Even in extremely thin geometries, crack roughening is a phenomenon that occupies the volume of the system. On the other hand, in hierarchical systems crack propagation is localized at the surface even if the surface is not weaker than the bulk. Remember the Gecko foot: The dilemma of reversible adhesion arises from the problem that strengthening interfacial adhesion locally may cause failure to occur elsewhere. Hierarchical architecture provides a work-around as it is able to localize detachment at the interface even if the interfacial adhesion forces are as strong as the cohesive forces in the bulk of the system. While the Gecko foot adheres strongly, it can still be pulled off without ripping the foot apart. 

\subsection*{Fracture strength}

The differences in fracture mechanisms between our two test systems suggest important practical implications also in the case in which a weak layer is introduced at the interface, which in our simulations is done by imposing stronger cohesive forces in the bulk ($t_1=10\,t_0$, i.e. $m=10$). In that case, the strength heterogeneity localizes damage at the interface and the crack becomes smooth even in case of non-hierarchical systems. However, this implies that these systems partly lose their ability to store damage and turn significantly weaker, whereas hierarchical systems might be only mildly affected. In order to verify this hypothesis, we run comparative simulations for systems with $m=1$ and $m=10$. In addition to systems with initially intact interface, we consider systems which contain pre-existing interface cracks of different length $a$. These cracks are introduced as flat ribbons of zero strength spanning the entire system in the $x$ direction, and extending over a length $a$ in the $y$ direction. Strength is evaluated by measuring the peak stress $\sigma_\mathrm{p}$ and the specific work of failure $w_\mathrm{f}=(1-a/L)^{-1}W_\mathrm{f}$, i.e., the area under the stress-strain curve normalized by the initially intact surface area fraction $(1-a/L)$. Both $\sigma_\mathrm{p}$ and $w_\mathrm{f}$ are averaged over multiple network realizations. 

\begin{figure}[!tbh]
\centering
\includegraphics[width=0.8\linewidth]{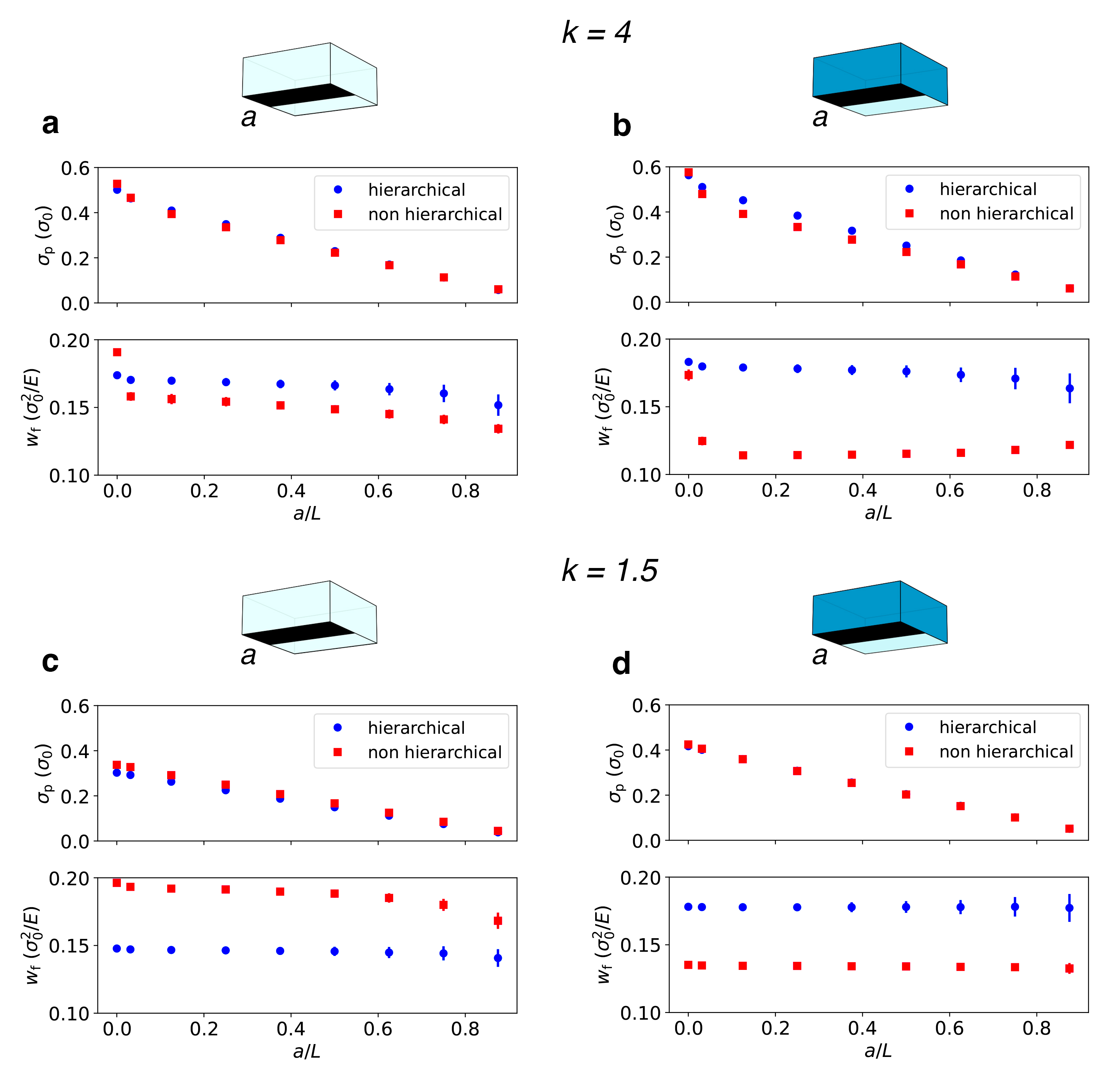}
\caption{Top: Fracture strength, peak stress and specific work of failure for varying notch sizes $a$; top: $k=4$, (a) $m=1$, (b) $m=10$; bottom: $k=1.5$, (c) $m=1$, (d) $m=10$.}
\label{fig:strength}
\end{figure}

Let us first discuss the results for $k=4$ (weak fluctuations in local strength, Figure \ref{fig:strength}, top). In the weak-interface case (Figure \ref{fig:strength}b, the hierarchical systems outperform the non-hierarchical ones slightly in terms of peak load and, prominently, in terms of work of failure. This behavior partly extends also to homogeneous systems (\ref{fig:strength}a), suggesting that the ability of hierarchical systems to localize damage comes at no significant cost in terms of strength performance. We remark that our weak-interface model is implemented by strengthening the connections in the bulk. For the hierarchical system this significantly increases the work of failure, whereas for the non-hierarchical system the work of failure decreases significantly as can no longer grow a rough crack surface (compare Figure \ref{fig:strength}, panels a and b).

We can now move to the results for $k=1.5$ (large fluctuations in local strength, Figure \ref{fig:strength}, bottom). Such systems show a lower peak stress than their less disordered counterparts, a finding which is long known in non hierarchical interface systems (see e.g. \cite{fyffe2004effects}). Hierarchical architecture does not significantly change this behavior, which changes only slightly when strengthening the bulk material. However, the specific work of fracture is strongly affected by the degree of disorder: In highly disordered systems with homogeneous mean strength, the specific work of failure is significantly increased for non hierarchical systems, whereas it is decreased for hierarchical ones, to the extent that these become actually weaker than their non hierarchical counterparts. In heterogeneous systems with weak interfaces, on the other hand, disorder has only a weak influence on the work of failure, which remains higher for hierarchical systems. 

To understand these findings we note that in bulk systems, large strength fluctuations tend to increase crack roughness, and extreme fluctuations may even give rise to a percolation-like failure scenario with super-rough fracture surfaces. Evidently, the idea that high disorder leads to enhanced roughening is consistent with the increase of work-of-failure in non hierarchical systems of homogeneous strength. In systems with large strength discrepancies between bulk and interface, on the other hand, the work of failure remains unchanged as the fracture surface remains essentially flat. The same argument makes it plausible that the work of failure does not change with disorder for the hierarchical systems, where the fracture surface is localized at the interface irrespective of the degree of disorder. However, the above argument still offers no clear cut explanation why the work of failure in hierarchical systems is actually increased when we increase strength in the bulk. 

\subsection*{Multiscaling analysis and correlations}

\begin{figure}[t]
\centering
\includegraphics[width=0.7\linewidth]{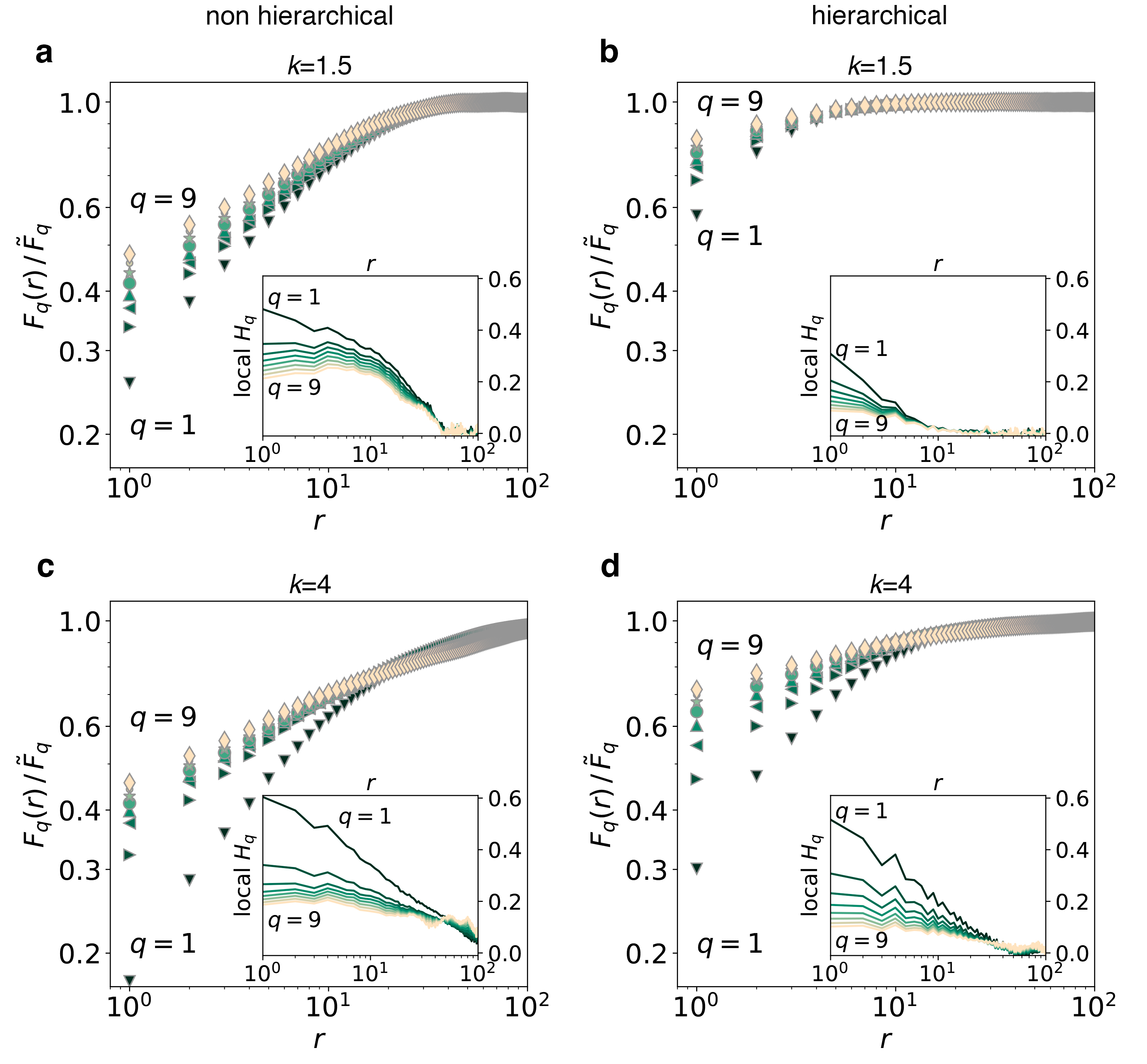}
\caption{Multiscaling analysis of crack surfaces. (a) Non-hierarchical system, $k=1.5$. 
(b) Hierarchical system, $k=1.5$.   
(c) Non-hierarchical system, $k=4$.  
(d) Hierarchical system, $k=4$. Integer values of $q$ between $1$ and $9$ are shown (from darker to brighter symbols). Insets show estimates of the local values of the exponent $H_q$. Systems of size $L=256$ and thickness $L_z=9$ are considered.}
\label{fig:multiscaling}
\end{figure}

We expect the different fracture modes of our systems to leave measurable traces in the statistical properties of their fracture surfaces. Figure \ref{fig:multiscaling} shows results of multiscaling analysis of fracture surfaces from non-hierarchical and hierarchical systems, for strong and weak fluctuations in local strengths  (see Methods section). Multiscaling is observed in all cases, as clearly indicated by the $q$ dependent   $F_q(r)$ curves and the plots of the local values of $H_q$.

Evidently, the power law regimes must be truncated as the height $h$ is constrained by the film height. This simple fact makes it unlikely to see scaling regimes of more than one order of magnitude in height difference (the elementary surface step is $1$ lattice unit, and the maximum height difference is $L_z-1$, that is $8$ lattice units for the largest $L=256$ system we consider). For hierarchical systems, however, the saturation of the structure function occurs earlier than for non hierarchical ones.
We note here that the quantities $F_q(r)$ carry information about both height fluctuations and height correlations in $h$, with plateaus corresponding to regions of highly suppressed correlations. In the hierarchical systems, such plateaus  indicating loss of surface correlations appear already at distances $r\approx L_z=9$. 

\begin{figure}[t]
\centering
\includegraphics[width=0.7\linewidth]{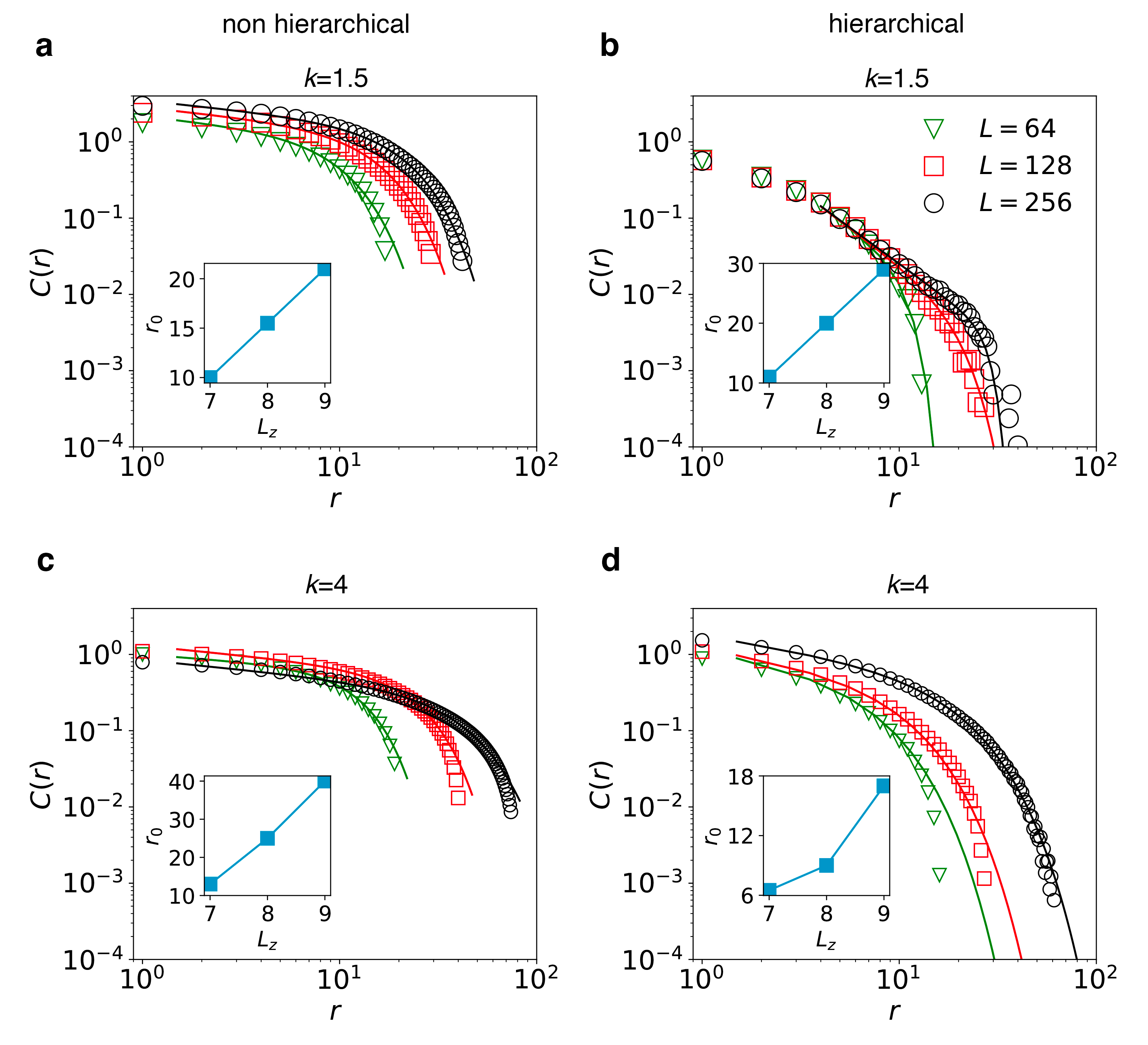}
\caption{Correlation analysis of crack surfaces. (a) Non-hierarchical system, $k=1.5$.   
(b) Hierarchical system, $k=1.5$.   
(c) Non-hierarchical system, $k=4$.  
(d) Hierarchical system, $k=4$.
Truncated power-laws are used to fit the data, with cutoff distances $r_0$ plotted in the insets, and exponents: 
(a) $\eta=0.26$, $\zeta=1.8$,
(b) $\eta=1.9$, $\zeta=5$,
(c) $\eta=0.25$, $\zeta=2.2,\,2,\,1.6$,
(d) $\eta=0.4$, $\zeta=1.35$.
}
\label{fig:correlations}
\end{figure}

To clarify the role of correlations alone we chose to measure the correlation function $C(r)$, performing a detailed size-scaling analysis (see Methods section). Results for $C(r)$ are collected in Figure \ref{fig:correlations}. Correlation functions exhibit truncated power-law dependence on the distance $r$ of the form
\begin{equation}\label{eq:truncated}
C(r)\approx r^{-\eta} \exp\left[ -\left(\frac{r}{r_0}\right)^\zeta \right]
\end{equation}
While the very limited range of $r$ makes the estimate of the exponents $\eta$ and $\zeta$ imprecise, by fitting the data we clearly confirm that the origin of the short-distance cut-off $r_0$ is the systems thickness $L_z$, with an approximately linear relationship between the two quantities in almost all cases. The actual correlation values, however, differ dramatically: for $r\gg L_z$ hierarchical systems display significantly lower correlations than those observed in non-hierarchical systems, even by two orders of magnitude in the case of $k=1.5$.

While in both systems height correlations along the $xy$ plane are screened at distances beyond $L_z$, the gap structure of hierarchical systems may further weaken residual correlations. 
Interestingly, variations in $k$ do not seem to heavily affect the relationship between $r_0$ and $L_z$, suggesting that, even though fracture patterns are a result of the interplay of elastic stress redistribution and local failure, even elasticity alone might help explain not just the rather intuitive screening effect of the system's thickness, but also the suppression of height correlations in hierarchical systems, as we show in the following.

\subsection*{Green's functions}
In order to study the mechanisms of stress redistribution in our thin film geometries, and in an attempt to clarify the differences in screening mechanisms, we compute the Green's function of the equilibrium equation \ref{eq:discrete_laplace}. We note that the equilibrium equation is a discrete Poisson equation, which admits discrete Green's functions that depend on the network structure (see Methods). Before computing those, let us consider the continuum limit of the problem, i.e. an isotropic box of sizes $L\times L\times L_z$. A series representation of the Green's function $G(\mathbf{r},\mathbf{r'})$  is well known from electrostatics \cite{Jackson_book}, however the convergence of the series is slow, making it hard to truncate sums and highlight the thickness dependence in a thin-film geometry \cite{Taloni2015_SciRep}. Here we choose to compute a different representation of  $G(\mathbf{r}, \mathbf{r'})$, which makes the role of $L_z$ more evident, and the transition to the network formalism more natural, using Fredholm theory (see Methods). Equation \ref{eq:green_solution} is our exact result for the elastic Green's function, for the scalar problem, in a continuous isotropic medium, and in the $L\to\infty$ limit. Its asymptotic expansion in Equation \ref{eq:first_order} shows how elastic load redistribution is screened for distances larger than $L_z$. In regard to our network models, it should approximate the behavior of the non-hierarchical systems, where quenched impurities are isotropically distributed on average, but not that of the hierarchical system. 

\begin{figure}[ht]
\centering
\includegraphics[width=0.8\linewidth]{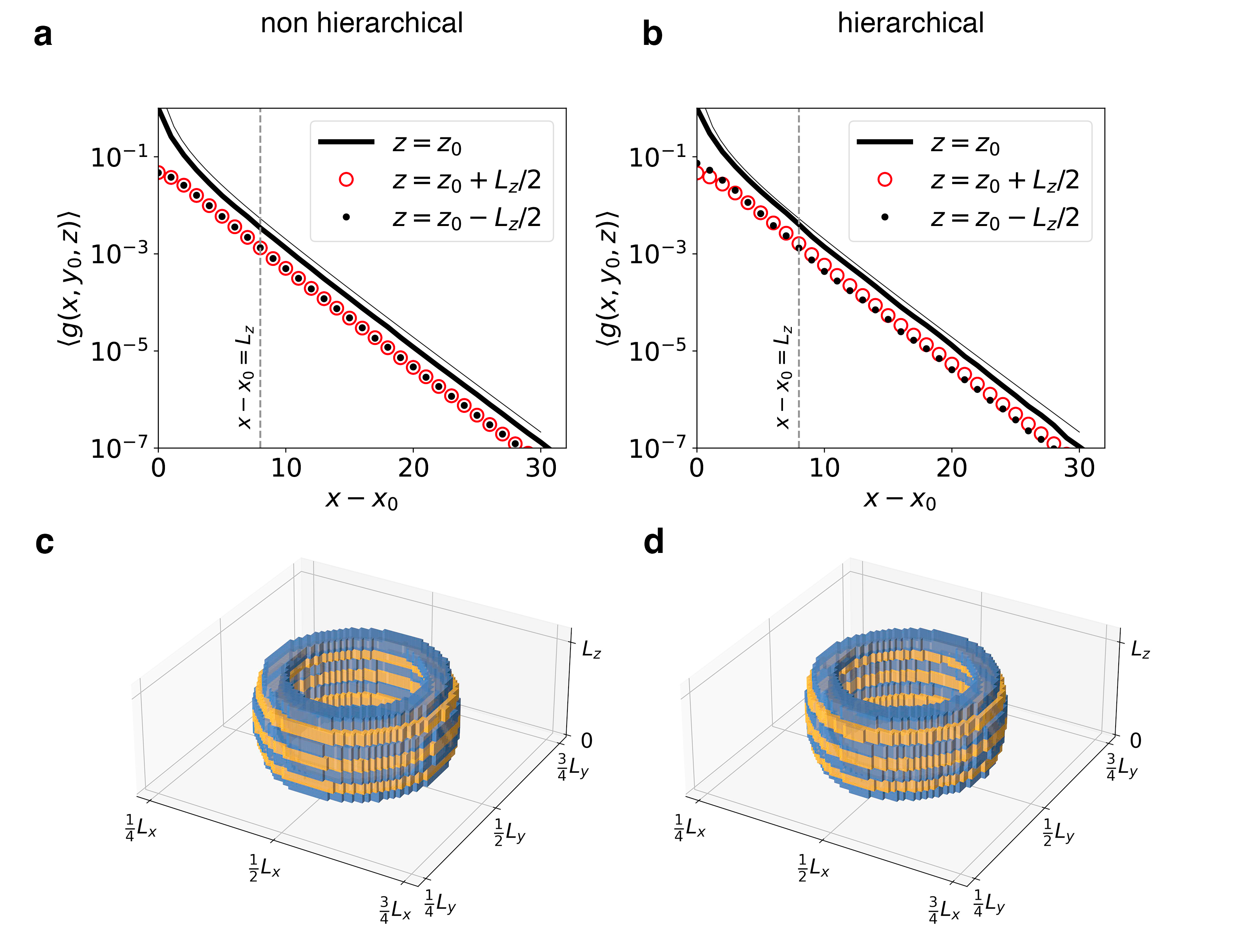}
\caption{Average network Green's functions.
A point perturbation is located in the center, $(x_0,y_0,z_0)=(L,L,L_z)/2$. Green's functions are averaged over multiple network realizations. Top row: radial Green's function, computed along the $x$ direction, at three different heights, for non-hierarchical (a) and hierarchical systems (b). Thin solid lines represent the Green's function of the continuum isotropic problem for $z=z_0$, obtained evaluating the first $10^5$ terms of the sum in Equation \ref{eq:green_solution}, and rescaled in height in order to allow for direct comparison with the corresponding discrete Green's function (thick solid lines). Bottom row: average equipotential surfaces for non-hierarchical (c) and hierarchical (d) systems, with values of $G\in[10^{-6},10^{-5}]$. Alternating colors are used to distinguish contiguous values of $z$.}
\label{fig:green_average}
\end{figure}

Next, we transfer the Green's function formalism to the discrete case. In the Methods section we introduced the discrete Green's function $\mathbb{G}$ of our problem, as the inverse of a non-singular submatrix of the graph Laplacian matrix $\mathbb{L}$. Our goal is now to extract information from $\mathbb{G}$, which can be directly compared to the continuum Green function in Equations \ref{eq:green_solution} and \ref{eq:first_order}. We noted in particular that $G_{ij}$ is the displacement at node $i$ of coordinates $(x,y,z)$, as induced by a unit force acting at node $j$ of coordinates $(x_0,y_0,z_0)$, which allows us to identify $G_{ij}$ ans the discrete counterpart of $G(x,y,z;x_0,y_0,z_0)$.
In particular we choose as $j=j_0$ the node at the center of the network, $(x_0,y_0,z_0)=(L,L,L_z)/2$. In Figure \ref{fig:green_average} we plot the ensemble averaged Green function $\langle g(x,y,z)\rangle$, normalized as $g(x,y,z)=G_{ij_0}/G_{j_0j_0}$ and averaged over many network realizations. Let us first discuss the radial decay of $\langle g(x,y,z)\rangle$, which based on the continuum prediction we expect to show an exponential tail at distances larger than $L_s$. Both the non hierarchical and the hierarchical system match this behavior nearly perfectly. 3D plots show that both for hierarchical and non hierarchical systems the ensemble averaged Greens functions are isotropic and both their radial and angular behavior is, in fact, nearly identical. This observation is, of course, astonishing in view of the significant differences in average fracture behavior.

\begin{figure}[ht]
\centering
\includegraphics[width=0.8\linewidth]{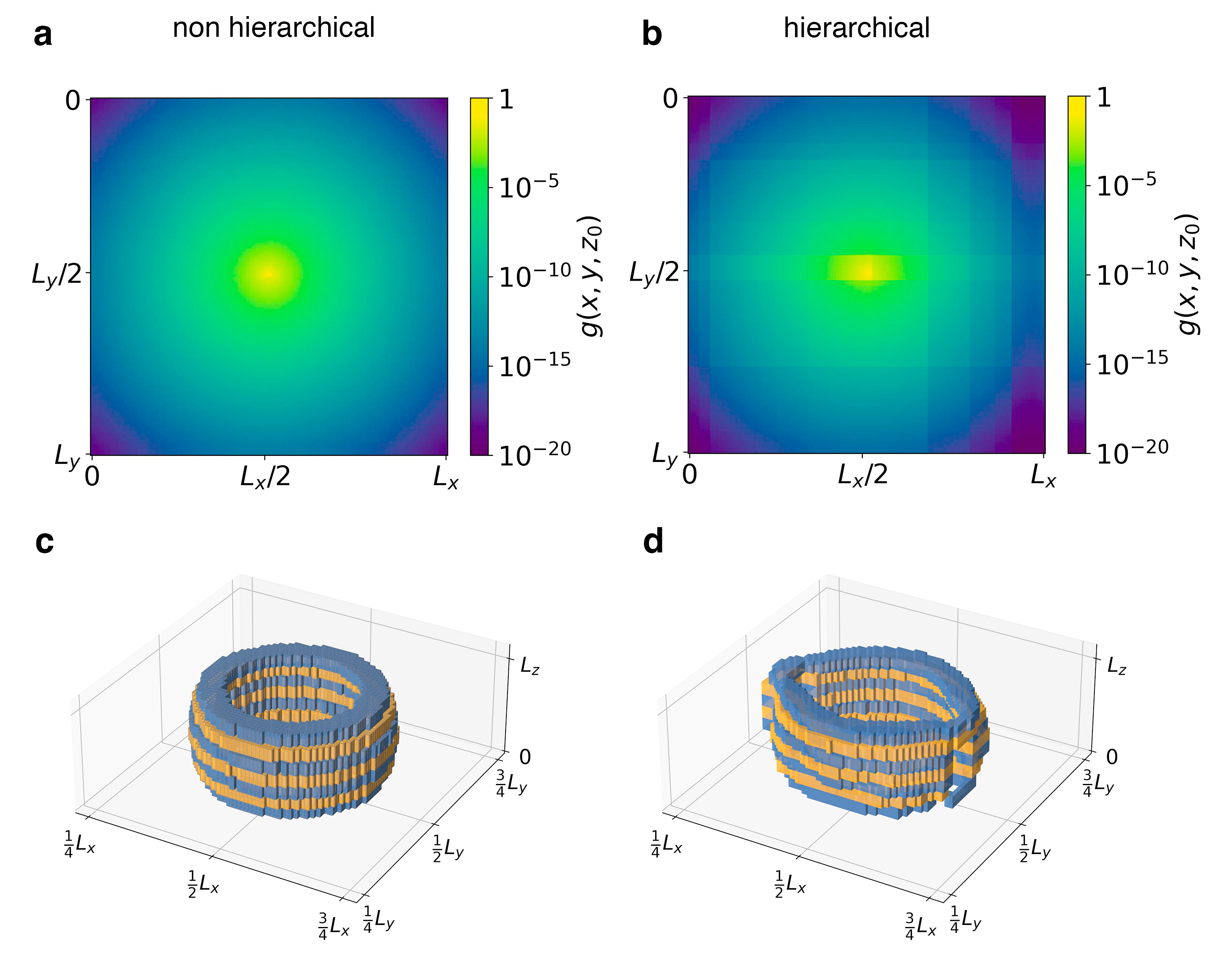}
\caption{Network Green's functions for single realizations, $L=128$. A point perturbation is located in the center, $(x_0,y_0,z_0)=(L,L,L_z)/2$. Green's functions are evaluated for a non-hierarchical system (a) and a hierarchical system (b), at height $z_0$ and across the $xy$ plane. Variation along the $z$ direction is shown in the 3D plots, for non-hierarchical (c) and hierarchical (d) systems, as equipotential surfaces with values of $G\in[10^{-6},10^{-5}]$. Alternating colors are used to distinguish contiguous values of $z$.  }
\label{fig:green_single}
\end{figure}

To resolve this apparent paradox, we observe that fracture is by nature a strongly nonlinear process: our simulations study failure of individual realizations, not of ensemble averages. We therefore ask   how the network Green's functions of single realizations  as shown in Figure \ref{fig:green_single} differ from their ensemble averaged counterparts. For a non hierarchical network, there is not much difference: The network Green's function still exhibits an approximately monotonic decay and polar symmetry in the $xy$ plane. However, the Green's function of the hierarchical system differs signficantly from the ensemble average: It displays a pronounced anisotropy as well as localized discontinuities corresponding to the gaps introduced in the hierarchical system. 
These features are direct reflections of the cut patterns of the hierarchical network: A cut introduces local discontinuities into the Green's function as it locally interrupts stress transmission, and it introduces local anisotropies because this interruption affects only the direction perpendicular to the cut. As a consequence, cracks propagating towards the cuts may be arrested and deflected sideways. These observations indicate that the quenched anisotropy in the single-network hierarchical Green function is not an undesired form of noise or fluctuation, it is in fact the main driver of the crack arrest phenomenon. Computing averages only masks the essential anisotropy and discontinuity of the Green's function on the single realization level. Finally we note that Figures \ref{fig:green_average} and \ref{fig:green_single} are independent of the choice of the Weibull shape parameter $k$, as they are obtained for the initial network states.

\section*{Discussion}
Our work is motivated by the possibility of understanding and controlling the fracture response of certain architected materials by imposing specific microstructural patterns. If microstructure can influence fracture toughness and inessential work of failure in the bulk, can it also be harnessed to prescribe failure locations? And if so, does this ability come at a price in terms of materials strength? Our results show that simple hierarchical microstructures, inspired by smart biological adhesion systems, can indeed control detachment locations, by localizing the fracture process at a prescribed interface. At the same time these microstructures impede crack propagation, by mitigating stress concentrations at crack tips. This form of interface confinement is substantially different form the fracture behavior of a statistically isotropic heterogeneous thin film, whose fracture strength depends on volumetric contributions, namely the screened stress redistribution and more importantly the ability to store damage throughout the entire volume. 

The interface confinement property of hierarchical systems becomes advantageous in problems of adhesion, where cohesive forces exceed adhesive forces. In such cases, the strength differential between bulk and interface always localizes damage at the interface. As a consequence, non-hierarchical systems cannot rely on their ability to store damage in terms of fracture surface roughening and become inherently weak. Hierarchical systems, instead, are not strongly affected by stronger cohesion in the bulk, exactly because their fracture behavior always forces damage localization at the interface, whereas their strength relies on gaps interrupting stress transmission and favoring the arrest of cracks before they become critical. 
The opposite limit, where cohesive and adhesive forces are comparable, is also interesting as it allows us to further investigate the role of microstructure in the failure mechanism. In this case, the fracture behavior of non-hierarchical system is the same that we would expect for a statistically isotropic disordered system, where correlations are screened beyond a distance equal to $L_z$. While we are able to verify that the same screening distance controls the response of hierarchical samples, we find that correlations in fracture surfaces are heavily suppressed, in agreement with the observation of crack arrest in hierarchical materials. 

While it can be loosely argued that hierarchical structures achieve crack arrest by eliminating stress concentrations, our theoretical formalism allows us to more precisely conclude that the anisotropy of the elastic kernel, which only becomes manifest in single realizations, is at the root of crack arrest, as it redirects stresses away from planar defects that intercept a propagating crack. Redirection of stresses most of the times delays crack growth by forcing the system to nucleate new microcracks over and over again, but this is by no means a universal principle and Figure \ref{fig:strength}c shows one such exception. Computing the network Green's function $G_{ij}$ at a precise node $j$, as we did here, provides information about the elastic response due to a perturbation in that specific location of the network. Future studies can now focus on a more general approach consisting in the analysis of the leading terms of its spectral resolution
$
\mathrm{G}=\sum_{n\in\Omega\setminus\partial\Omega}(1/\lambda_n)\mathbf{v_n}\otimes\mathbf{v_n},
$
i.e. the discrete version of Equation \ref{eq:fredholm}, where $\mathbf{v_n}$ is the eigenvector associated with the eigenvalue $1/\lambda_n$. In particular, the study of the local properties of the projection operators $\mathbf{v_n}\otimes\mathbf{v_n}$ should allow to harness network localization properties to control the anisotropy of $G$, for configurations not limited to the pristine initial state, and even for more complex structures than the idealized ones considered here.  We emphasize that this type of spectral analysis carries over to any type of discrete balance equation of the form
$
\mathbb{K}\mathbf{u}=\mathbf{f},
$ 
relating the vectors of generalized displacements $\mathbf{u}$ and momenta $\mathbf{f}$ through the stiffness matrix $\mathbb{K}$ of the network, where $\mathbb{G}=\mathbb{K}^{-1}$. Our simulation results were obtained under the scalar elasticity assumption of the RFM. The RFM correctly reproduces fundamental results of fracture mechanics, such as the stress concentration at the crack tip, and becomes exact in limiting cases such as that of anti-plane shear deformation and that of materials with vanishing Poisson ratio. While it remains an effective approximation, especially in the case of uni-axial tension, we believe that more accurate beam-model simulations are needed in the case of, e.g., shear deformation. As noted above, even in the case of beam simulations, our Green's function formalism will apply. 

Graph theoretical approaches such as the one developed here prove especially useful in detecting localized patterns of deformation and damage, both in numerical studies and in experiments. In particular, in systems where microstucture prevents load accumulation and crack growth, localization measurements may prove useful in interpreting digital image correlation (DIC)  data \cite{Mustalahti2010,Miksic2011}. Similarly, a focus on network based microstructure description paves the way to  alternative approaches consisting in the use of machine learning techniques. Machine learning can be applied to predict failure properties of metamaterials \cite{hiemer2022predicting}, or to optimize the design of ordered and disordered hierarchical metamaterials in view of enhancing their mechanical performance \cite{yu2022multiresolution,zaiser2023disordered,luu2023,buehler2023msm}.
More generally, similar data-centric approaches may prove essential to optimize automated design strategies for complex and architected functional materials, an important task for future investigations.

\bibliography{adhesion}

\begin{thebibliography}{10}
\urlstyle{rm}
\expandafter\ifx\csname url\endcsname\relax
  \def\url#1{\texttt{#1}}\fi
\expandafter\ifx\csname urlprefix\endcsname\relax\def\urlprefix{URL }\fi
\expandafter\ifx\csname doiprefix\endcsname\relax\def\doiprefix{DOI: }\fi
\providecommand{\bibinfo}[2]{#2}
\providecommand{\eprint}[2][]{\url{#2}}

\bibitem{Bonamy2006_PRL}
\bibinfo{author}{Bonamy, D.}, \bibinfo{author}{Ponson, L.},
  \bibinfo{author}{Prades, S.}, \bibinfo{author}{Bouchaud, E.} \&
  \bibinfo{author}{Guillot, C.}
\newblock \bibinfo{journal}{\bibinfo{title}{Scaling exponents for fracture
  surfaces in homogeneous glass and glassy ceramics}}.
\newblock {\emph{\JournalTitle{Physical Review Letters}}}
  \textbf{\bibinfo{volume}{97}}, \bibinfo{pages}{135504}
  (\bibinfo{year}{2006}).

\bibitem{alava2006statistical}
\bibinfo{author}{Alava, M.~J.}, \bibinfo{author}{Nukala, P.~K.} \&
  \bibinfo{author}{Zapperi, S.}
\newblock \bibinfo{journal}{\bibinfo{title}{Statistical models of fracture}}.
\newblock {\emph{\JournalTitle{Advances in Physics}}}
  \textbf{\bibinfo{volume}{55}}, \bibinfo{pages}{349--476}
  (\bibinfo{year}{2006}).

\bibitem{alava2008role}
\bibinfo{author}{Alava, M.~J.}, \bibinfo{author}{Nukala, P.~K.} \&
  \bibinfo{author}{Zapperi, S.}
\newblock \bibinfo{journal}{\bibinfo{title}{Role of disorder in the size
  scaling of material strength}}.
\newblock {\emph{\JournalTitle{Physical review letters}}}
  \textbf{\bibinfo{volume}{100}}, \bibinfo{pages}{055502}
  (\bibinfo{year}{2008}).

\bibitem{Lakes1993}
\bibinfo{author}{Lakes, R.}
\newblock \bibinfo{journal}{\bibinfo{title}{Materials with structural
  hierarchy}}.
\newblock {\emph{\JournalTitle{Nature}}} \textbf{\bibinfo{volume}{361}},
  \bibinfo{pages}{511--515} (\bibinfo{year}{1993}).

\bibitem{fratzl2007nature}
\bibinfo{author}{Fratzl, P.} \& \bibinfo{author}{Weinkamer, R.}
\newblock \bibinfo{journal}{\bibinfo{title}{Nature’s hierarchical
  materials}}.
\newblock {\emph{\JournalTitle{Progress in materials Science}}}
  \textbf{\bibinfo{volume}{52}}, \bibinfo{pages}{1263--1334}
  (\bibinfo{year}{2007}).

\bibitem{gao2006application}
\bibinfo{author}{Gao, H.}
\newblock \bibinfo{title}{Application of fracture mechanics concepts to
  hierarchical biomechanics of bone and bone-like materials}.
\newblock In \emph{\bibinfo{booktitle}{Advances in Fracture Research: Honour
  and Plenary Lectures Presented at the 11 th International Conference on
  Fracture (ICF11), Held in Turin, Italy, on March 20--25, 2005}},
  \bibinfo{pages}{101--137} (\bibinfo{organization}{Springer},
  \bibinfo{year}{2006}).

\bibitem{Sen2011}
\bibinfo{author}{Sen, D.} \& \bibinfo{author}{Buehler, M.~J.}
\newblock \bibinfo{journal}{\bibinfo{title}{Structural hierarchies define
  toughness and defect-tolerance despite simple and mechanically inferior
  brittle building blocks}}.
\newblock {\emph{\JournalTitle{Scientific reports}}}
  \textbf{\bibinfo{volume}{1}}, \bibinfo{pages}{1--9} (\bibinfo{year}{2011}).

\bibitem{mirzaeifar2015defect}
\bibinfo{author}{Mirzaeifar, R.}, \bibinfo{author}{Dimas, L.~S.},
  \bibinfo{author}{Qin, Z.} \& \bibinfo{author}{Buehler, M.~J.}
\newblock \bibinfo{journal}{\bibinfo{title}{Defect-tolerant bioinspired
  hierarchical composites: simulation and experiment}}.
\newblock {\emph{\JournalTitle{ACS Biomaterials Science \& Engineering}}}
  \textbf{\bibinfo{volume}{1}}, \bibinfo{pages}{295--304}
  (\bibinfo{year}{2015}).

\bibitem{Moretti2018}
\bibinfo{author}{Moretti, P.}, \bibinfo{author}{Dietemann, B.},
  \bibinfo{author}{Esfandiary, N.} \& \bibinfo{author}{Zaiser, M.}
\newblock \bibinfo{journal}{\bibinfo{title}{Avalanche precursors of failure in
  hierarchical fuse networks}}.
\newblock {\emph{\JournalTitle{Scientific reports}}}
  \textbf{\bibinfo{volume}{8}}, \bibinfo{pages}{1--7} (\bibinfo{year}{2018}).

\bibitem{Sun2012}
\bibinfo{author}{Sun, J.} \& \bibinfo{author}{Bhushan, B.}
\newblock \bibinfo{journal}{\bibinfo{title}{Hierarchical structure and
  mechanical properties of nacre: a review}}.
\newblock {\emph{\JournalTitle{Rsc Advances}}} \textbf{\bibinfo{volume}{2}},
  \bibinfo{pages}{7617--7632} (\bibinfo{year}{2012}).

\bibitem{Jiao2015}
\bibinfo{author}{Jiao, D.}, \bibinfo{author}{Liu, Z.}, \bibinfo{author}{Zhang,
  Z.} \& \bibinfo{author}{Zhang, Z.}
\newblock \bibinfo{journal}{\bibinfo{title}{Intrinsic hierarchical structural
  imperfections in a natural ceramic of bivalve shell with distinctly graded
  properties}}.
\newblock {\emph{\JournalTitle{Scientific Reports}}}
  \textbf{\bibinfo{volume}{5}}, \bibinfo{pages}{1--13} (\bibinfo{year}{2015}).

\bibitem{Gao2006}
\bibinfo{author}{Gao, H.}
\newblock \bibinfo{journal}{\bibinfo{title}{Application of fracture mechanics
  concepts to hierarchical biomechanics of bone and bone-like materials}}.
\newblock {\emph{\JournalTitle{International Journal of fracture}}}
  \textbf{\bibinfo{volume}{138}}, \bibinfo{pages}{101--137}
  (\bibinfo{year}{2006}).

\bibitem{Rho1998}
\bibinfo{author}{Rho, J.-Y.}, \bibinfo{author}{Kuhn-Spearing, L.} \&
  \bibinfo{author}{Zioupos, P.}
\newblock \bibinfo{journal}{\bibinfo{title}{Mechanical properties and the
  hierarchical structure of bone}}.
\newblock {\emph{\JournalTitle{Medical engineering \& physics}}}
  \textbf{\bibinfo{volume}{20}}, \bibinfo{pages}{92--102}
  (\bibinfo{year}{1998}).

\bibitem{Gautieri2011}
\bibinfo{author}{Gautieri, A.}, \bibinfo{author}{Vesentini, S.},
  \bibinfo{author}{Redaelli, A.} \& \bibinfo{author}{Buehler, M.~J.}
\newblock \bibinfo{journal}{\bibinfo{title}{Hierarchical structure and
  nanomechanics of collagen microfibrils from the atomistic scale up}}.
\newblock {\emph{\JournalTitle{Nano letters}}} \textbf{\bibinfo{volume}{11}},
  \bibinfo{pages}{757--766} (\bibinfo{year}{2011}).

\bibitem{roemer2008prion}
\bibinfo{author}{L., R.} \& \bibinfo{author}{Scheibel, T.}
\newblock \bibinfo{journal}{\bibinfo{title}{The elaborate structure of spider
  silk: structure and function of a natural high performance fiber}}.
\newblock {\emph{\JournalTitle{Prion}}} \textbf{\bibinfo{volume}{2}},
  \bibinfo{pages}{154--161} (\bibinfo{year}{2008}).

\bibitem{lu2023silk}
\bibinfo{author}{Lu, W.}, \bibinfo{author}{Kaplan, D.~L.} \&
  \bibinfo{author}{Buehler, M.~J.}
\newblock \bibinfo{journal}{\bibinfo{title}{Generative modeling, design, and
  analysis of spider silk protein sequences for enhanced mechanical
  properties}}.
\newblock {\emph{\JournalTitle{Advanced Functional Materials}}}
  \bibinfo{pages}{2311324} (\bibinfo{year}{2023}).

\bibitem{Zaiser2022}
\bibinfo{author}{Zaiser, M.} \emph{et~al.}
\newblock \bibinfo{journal}{\bibinfo{title}{Hierarchical slice patterns inhibit
  crack propagation in brittle sheets}}.
\newblock {\emph{\JournalTitle{Physical Review Applied}}}
  \textbf{\bibinfo{volume}{18}}, \bibinfo{pages}{044035}
  (\bibinfo{year}{2022}).

\bibitem{pournajar2023failure}
\bibinfo{author}{Pournajar, M.} \emph{et~al.}
\newblock \bibinfo{journal}{\bibinfo{title}{Failure precursors and failure
  mechanisms in hierarchically patterned paper sheets in tensile and creep
  loading}}.
\newblock {\emph{\JournalTitle{Physical Review Applied}}}
  \bibinfo{pages}{024008} (\bibinfo{year}{2023}).

\bibitem{Hosseini2021}
\bibinfo{author}{Hosseini, S.~A.}, \bibinfo{author}{Moretti, P.},
  \bibinfo{author}{Konstantinidis, D.} \& \bibinfo{author}{Zaiser, M.}
\newblock \bibinfo{journal}{\bibinfo{title}{Beam network model for fracture of
  materials with hierarchical microstructure}}.
\newblock {\emph{\JournalTitle{International Journal of Fracture}}}
  \textbf{\bibinfo{volume}{227}}, \bibinfo{pages}{243--257}
  (\bibinfo{year}{2021}).

\bibitem{hosseini2023enhanced}
\bibinfo{author}{Hosseini, S.~A.}, \bibinfo{author}{Moretti, P.} \&
  \bibinfo{author}{Zaiser, M.}
\newblock \bibinfo{journal}{\bibinfo{title}{Enhanced fault tolerance in
  biomimetic hierarchical materials: A simulation study}}.
\newblock {\emph{\JournalTitle{Physical Review Materials}}}
  \textbf{\bibinfo{volume}{7}}, \bibinfo{pages}{053612} (\bibinfo{year}{2023}).

\bibitem{Puglisi2013_PRE}
\bibinfo{author}{Puglisi, G.} \& \bibinfo{author}{Truskinovsky, L.}
\newblock \bibinfo{journal}{\bibinfo{title}{Cohesion-decohesion asymmetry in
  geckos}}.
\newblock {\emph{\JournalTitle{Phys. Rev. E}}} \textbf{\bibinfo{volume}{87}},
  \bibinfo{pages}{032714} (\bibinfo{year}{2013}).

\bibitem{Esfandiary2022}
\bibinfo{author}{Esfandiary, N.}, \bibinfo{author}{Zaiser, M.} \&
  \bibinfo{author}{Moretti, P.}
\newblock \bibinfo{journal}{\bibinfo{title}{Statistical aspects of interface
  adhesion and detachment of hierarchically patterned structures}}.
\newblock {\emph{\JournalTitle{Journal of Statistical Mechanics: Theory and
  Experiment}}} \textbf{\bibinfo{volume}{2022}}, \bibinfo{pages}{023301}
  (\bibinfo{year}{2022}).

\bibitem{Costagliola2016_PRE}
\bibinfo{author}{Costagliola, G.}, \bibinfo{author}{Bosia, F.} \&
  \bibinfo{author}{Pugno, N.~M.}
\newblock \bibinfo{journal}{\bibinfo{title}{Static and dynamic friction of
  hierarchical surfaces}}.
\newblock {\emph{\JournalTitle{Phys. Rev. E}}} \textbf{\bibinfo{volume}{94}},
  \bibinfo{pages}{063003} (\bibinfo{year}{2016}).

\bibitem{Costagliola2022_IJSS}
\bibinfo{author}{Costagliola, G.}, \bibinfo{author}{Bosia, F.} \&
  \bibinfo{author}{Pugno, N.~M.}
\newblock \bibinfo{journal}{\bibinfo{title}{Correlation between slip precursors
  and topological length scales at the onset of frictional sliding}}.
\newblock {\emph{\JournalTitle{International Journal of Solids and
  Structures}}} \textbf{\bibinfo{volume}{243}}, \bibinfo{pages}{111525}
  (\bibinfo{year}{2022}).

\bibitem{KimJAST_2007}
\bibinfo{author}{Kim, T.~W.} \& \bibinfo{author}{Bhushan, B.}
\newblock \bibinfo{journal}{\bibinfo{title}{Adhesion analysis of multi-level
  hierarchical attachment system contacting with a rough surface}}.
\newblock {\emph{\JournalTitle{Journal of Adhesion Science and Technology}}}
  \textbf{\bibinfo{volume}{21}}, \bibinfo{pages}{1–20}
  (\bibinfo{year}{2007}).

\bibitem{Sauer2014_JAST}
\bibinfo{author}{Sauer, R.~A.}
\newblock \bibinfo{journal}{\bibinfo{title}{Advances in the computational
  modeling of the gecko adhesion mechanism}}.
\newblock {\emph{\JournalTitle{Journal of Adhesion Science and Technology}}}
  \textbf{\bibinfo{volume}{28}}, \bibinfo{pages}{240–255}
  (\bibinfo{year}{2014}).

\bibitem{Bhushan2009_PTRS}
\bibinfo{author}{Bhushan, B.}, \bibinfo{author}{Jung, Y.~C.} \&
  \bibinfo{author}{Koch, K.}
\newblock \bibinfo{journal}{\bibinfo{title}{Micro-, nano-and hierarchical
  structures for superhydrophobicity, self-cleaning and low adhesion}}.
\newblock {\emph{\JournalTitle{Philosophical Transactions of the Royal Society
  of London A: Mathematical, Physical and Engineering Sciences}}}
  \textbf{\bibinfo{volume}{367}}, \bibinfo{pages}{1631–1672}
  (\bibinfo{year}{2009}).

\bibitem{Gao2005_MM}
\bibinfo{author}{Gao, H.}, \bibinfo{author}{Wang, X.}, \bibinfo{author}{Yao,
  H.}, \bibinfo{author}{Gorb, S.} \& \bibinfo{author}{Arzt, E.}
\newblock \bibinfo{journal}{\bibinfo{title}{Mechanics of hierarchical adhesion
  structures of geckos}}.
\newblock {\emph{\JournalTitle{Mechanics of Materials}}}
  \textbf{\bibinfo{volume}{37}}, \bibinfo{pages}{275–285}
  (\bibinfo{year}{2005}).

\bibitem{fyffe2004effects}
\bibinfo{author}{Fyffe, B.} \& \bibinfo{author}{Zaiser, M.}
\newblock \bibinfo{journal}{\bibinfo{title}{The effects of snow variability on
  slab avalanche release}}.
\newblock {\emph{\JournalTitle{Cold regions science and technology}}}
  \textbf{\bibinfo{volume}{40}}, \bibinfo{pages}{229--242}
  (\bibinfo{year}{2004}).

\bibitem{Barai2013_PRE}
\bibinfo{author}{Barai, P.}, \bibinfo{author}{Nukala, P. K. V.~V.},
  \bibinfo{author}{Alava, M.~J.} \& \bibinfo{author}{Zapperi, S.}
\newblock \bibinfo{journal}{\bibinfo{title}{Role of the sample thickness in
  planar crack propagation}}.
\newblock {\emph{\JournalTitle{Phys. Rev. E}}} \textbf{\bibinfo{volume}{88}},
  \bibinfo{pages}{042411}, \doiprefix\url{10.1103/PhysRevE.88.042411}
  (\bibinfo{year}{2013}).

\bibitem{Taloni2015_SciRep}
\bibinfo{author}{Taloni, A.}, \bibinfo{author}{Benassi, A.},
  \bibinfo{author}{Sandfeld, S.} \& \bibinfo{author}{Zapperi, S.}
\newblock \bibinfo{journal}{\bibinfo{title}{Scalar model for frictional
  precursors dynamics}}.
\newblock {\emph{\JournalTitle{Scientific Reports}}}
  \textbf{\bibinfo{volume}{5}}, \bibinfo{pages}{8086} (\bibinfo{year}{2015}).

\bibitem{yao2024mechanical}
\bibinfo{author}{Yao, Z.}, \bibinfo{author}{Nasiri, S.}, \bibinfo{author}{Yang,
  M.} \& \bibinfo{author}{Zaiser, M.}
\newblock \bibinfo{journal}{\bibinfo{title}{Mechanical properties of interfaces
  between mg and sic: An ab initio study}}.
\newblock {\emph{\JournalTitle{Metals}}} \textbf{\bibinfo{volume}{14}},
  \bibinfo{pages}{467} (\bibinfo{year}{2024}).

\bibitem{deArcangelis1985_JPL}
\bibinfo{author}{{de Arcangelis}, L.}, \bibinfo{author}{Redner, S.} \&
  \bibinfo{author}{Herrmann, H.~J.}
\newblock \bibinfo{journal}{\bibinfo{title}{A random fuse model for breaking
  processes}}.
\newblock {\emph{\JournalTitle{J. Phys. Lett.}}} \textbf{\bibinfo{volume}{46}},
  \bibinfo{pages}{L585} (\bibinfo{year}{1985}).

\bibitem{chung2000_green}
\bibinfo{author}{Chung, F.} \& \bibinfo{author}{Yau, S.~T.}
\newblock \bibinfo{journal}{\bibinfo{title}{Discrete green's functions}}.
\newblock {\emph{\JournalTitle{Journal of Combinatorial Theory, Series A}}}
  \textbf{\bibinfo{volume}{91}}, \bibinfo{pages}{191--214},
  \doiprefix\url{https://doi.org/10.1006/jcta.2000.3094}
  (\bibinfo{year}{2000}).

\bibitem{Moretti2019_EPJB}
\bibinfo{author}{Moretti, P.}, \bibinfo{author}{Renner, J.},
  \bibinfo{author}{Safari, A.} \& \bibinfo{author}{Zaiser, M.}
\newblock \bibinfo{journal}{\bibinfo{title}{Graph theoretical approaches for
  the characterization of damage in hierarchical materials}}.
\newblock {\emph{\JournalTitle{Eur. J. Phys. B}}}
  \textbf{\bibinfo{volume}{92}}, \bibinfo{pages}{97} (\bibinfo{year}{2019}).

\bibitem{Jackson_book}
\bibinfo{author}{Jackson, J.~D.}
\newblock \emph{\bibinfo{title}{{Classical electrodynamics; 2nd ed.}}}
  (\bibinfo{publisher}{Wiley}, \bibinfo{address}{New York, NY},
  \bibinfo{year}{1975}).

\bibitem{Barabasi1991_PRA}
\bibinfo{author}{Barab\'asi, A.-L.} \& \bibinfo{author}{Vicsek}.
\newblock \bibinfo{journal}{\bibinfo{title}{Multifractality of self-affine
  fractals}}.
\newblock {\emph{\JournalTitle{Phys. Rev. A}}} \textbf{\bibinfo{volume}{44}},
  \bibinfo{pages}{2730--2733} (\bibinfo{year}{1991}).

\bibitem{Barabasi1992_PRA}
\bibinfo{author}{Barab\'asi, A.-L.} \emph{et~al.}
\newblock \bibinfo{journal}{\bibinfo{title}{Multifractality of growing
  surfaces}}.
\newblock {\emph{\JournalTitle{Phys. Rev. A}}} \textbf{\bibinfo{volume}{45}},
  \bibinfo{pages}{R6951--R6954} (\bibinfo{year}{1992}).

\bibitem{Picallo2009_PRL}
\bibinfo{author}{Picallo, C.~B.}, \bibinfo{author}{L{\'o}pez, J.~M.},
  \bibinfo{author}{Zapperi, S.} \& \bibinfo{author}{Alava, M.~J.}
\newblock \bibinfo{journal}{\bibinfo{title}{Optimization and plasticity in
  disordered media}}.
\newblock {\emph{\JournalTitle{Physical Review Letters}}}
  \textbf{\bibinfo{volume}{103}}, \bibinfo{pages}{225502}
  (\bibinfo{year}{2009}).

\bibitem{bustingorry2021_JoP}
\bibinfo{author}{Bustingorry, S.}, \bibinfo{author}{Guyonnet, J.},
  \bibinfo{author}{Paruch, P.} \& \bibinfo{author}{Agoritsas, E.}
\newblock \bibinfo{journal}{\bibinfo{title}{A numerical study of the statistics
  of roughness parameters for fluctuating interfaces}}.
\newblock {\emph{\JournalTitle{Journal of Physics: Condensed Matter}}}
  \textbf{\bibinfo{volume}{33}}, \bibinfo{pages}{345001}
  (\bibinfo{year}{2021}).

\bibitem{Mustalahti2010}
\bibinfo{author}{Mustalahti, M.}, \bibinfo{author}{Rosti, J.},
  \bibinfo{author}{Koivisto, J.} \& \bibinfo{author}{Alava, M.~J.}
\newblock \bibinfo{journal}{\bibinfo{title}{Relaxation of creep strain in
  paper}}.
\newblock {\emph{\JournalTitle{Journal of Statistical Mechanics: Theory and
  Experiment}}} \textbf{\bibinfo{volume}{2010}}, \bibinfo{pages}{P07019}
  (\bibinfo{year}{2010}).

\bibitem{Miksic2011}
\bibinfo{author}{Miksic, A.}, \bibinfo{author}{Koivisto, J.} \&
  \bibinfo{author}{Alava, M.}
\newblock \bibinfo{journal}{\bibinfo{title}{Statistical properties of low cycle
  fatigue in paper}}.
\newblock {\emph{\JournalTitle{Journal of Statistical Mechanics: Theory and
  Experiment}}} \textbf{\bibinfo{volume}{2011}}, \bibinfo{pages}{P05002}
  (\bibinfo{year}{2011}).

\bibitem{hiemer2022predicting}
\bibinfo{author}{Hiemer, S.}, \bibinfo{author}{Moretti, P.},
  \bibinfo{author}{Zapperi, S.} \& \bibinfo{author}{Zaiser, M.}
\newblock \bibinfo{journal}{\bibinfo{title}{Predicting creep failure by machine
  learning-which features matter?}}
\newblock {\emph{\JournalTitle{Forces in Mechanics}}}
  \textbf{\bibinfo{volume}{9}}, \bibinfo{pages}{100141} (\bibinfo{year}{2022}).

\bibitem{yu2022multiresolution}
\bibinfo{author}{Yu, C.-H.} \emph{et~al.}
\newblock \bibinfo{journal}{\bibinfo{title}{Hierarchical multiresolution design
  of bioinspired structural composites using progressive reinforcement
  learning}}.
\newblock {\emph{\JournalTitle{Advanced Theory and Simulations}}}
  \textbf{\bibinfo{volume}{5}}, \bibinfo{pages}{2200459}
  (\bibinfo{year}{2022}).

\bibitem{zaiser2023disordered}
\bibinfo{author}{Zaiser, M.} \& \bibinfo{author}{Zapperi, S.}
\newblock \bibinfo{journal}{\bibinfo{title}{Disordered mechanical
  metamaterials}}.
\newblock {\emph{\JournalTitle{Nature Reviews Physics}}}
  \textbf{\bibinfo{volume}{5}}, \bibinfo{pages}{679--688}
  (\bibinfo{year}{2023}).

\bibitem{luu2023}
\bibinfo{author}{Luu, R.~K.} \& \bibinfo{author}{Buehler, M.~J.}
\newblock \bibinfo{journal}{\bibinfo{title}{Bioinspiredllm: Conversational
  large language model for the mechanics of biological and bio-inspired
  materials}}.
\newblock {\emph{\JournalTitle{Advanced Science}}}  (\bibinfo{year}{2023}).

\bibitem{buehler2023msm}
\bibinfo{author}{Buehler, M.~J.}
\newblock \bibinfo{journal}{\bibinfo{title}{A computational building block
  approach towards multiscale architected materials analysis and design with
  application to hierarchical metal metamaterials}}.
\newblock {\emph{\JournalTitle{Modelling and Simulation in Materials Science
  and Engineering}}} \textbf{\bibinfo{volume}{31}}, \bibinfo{pages}{054001}
  (\bibinfo{year}{2023}).

\end{thebibliography}

\section*{Acknowledgements}

This research was funded by the Deutsche Forschungsgemeinschaft (DFG, German Research Foundation) - 377472739/GRK 2423/2-2023. The authors are very grateful for this support.

\section*{Author contributions statement}
C.G., P.M. and M.Z. devised the investigation.
C.G. performed the numerical simulations and analyzed the data. 
P.M. contributed the theoretical model.
All authors reviewed and approved the final version of the manuscript.

\section*{Additional information}

\textbf{Competing interests} The authors declare no competing interests.

\end{document}